%% file: sample-sigconf.tex
\documentclass[sigconf]{acmart}

\usepackage{amsmath,amssymb,amsfonts}
\usepackage{algorithmic}
\usepackage{graphicx}
\usepackage{textcomp}
\usepackage{xcolor}

\usepackage{tikz}
\usepackage{tabularx}
\usepackage{multirow}
\usepackage{array}
\usepackage{ifpdf}
\usepackage{url}
\usepackage{hyperref}
\hypersetup{hidelinks,
    colorlinks=true,
    allcolors=black,
    pdfstartview=Fit,
    breaklinks=true}
\usepackage{mdwlist}  
\usepackage{diagbox}

\usepackage{enumitem}

\newcommand{\ignore}[1]{}

\newtheorem{insight}{Insight}
\newtheorem{definition}{Definition}

\def\BibTeX{{\rm B\kern-.05em{\sc i\kern-.025em b}\kern-.08em
    T\kern-.1667em\lower.7ex\hbox{E}\kern-.125emX}}

\AtBeginDocument{%
  \providecommand\BibTeX{{%
    Bib\TeX}}}

\copyrightyear{2024} 
\acmYear{2024} 
\setcopyright{acmlicensed}\acmConference[ICSE '24]{2024 IEEE/ACM 46th International Conference on Software Engineering}{April 14--20, 2024}{Lisbon, Portugal}
\acmBooktitle{2024 IEEE/ACM 46th International Conference on Software Engineering (ICSE '24), April 14--20, 2024, Lisbon, Portugal}
\acmDOI{10.1145/3597503.3639218}
\acmISBN{979-8-4007-0217-4/24/04}

\acmPrice{15.00}

\begin{document}

\title{On the Effectiveness of Function-Level Vulnerability Detectors for Inter-Procedural Vulnerabilities}

\author{Zhen Li}
\authornote{National Engineering Research Center for Big Data Technology and System, Services Computing Technology and System Lab, Hubei Key Laboratory of Distributed System Security, Hubei Engineering Research Center on Big Data Security, Cluster and Grid Computing Lab}
\authornote{JinYinHu Laboratory, Wuhan, China}
\affiliation{%
 \institution{School of Cyber Science and Engineering, Huazhong University of Science and Technology}
 \city{Wuhan}
 \country{China}
 }
\email{zh\_li@hust.edu.cn}

\author{Ning Wang}
\authornotemark[1]
\affiliation{%
 \institution{School of Cyber Science and Engineering, Huazhong University of Science and Technology}
 \city{Wuhan}
 \country{China}
 }
\email{wangn@hust.edu.cn}

\author{Deqing Zou}
\authornotemark[1]
\authornotemark[2]
\authornote{Corresponding author}
\affiliation{%
  \institution{School of Cyber Science and Engineering, Huazhong University of Science and Technology}
  \city{Wuhan}
  \country{China}}
\email{deqingzou@hust.edu.cn}

\author{Yating Li}
\authornotemark[1]
\affiliation{%
 \institution{School of Cyber Science and Engineering, Huazhong University of Science and Technology}
 \city{Wuhan}
 \country{China}
 }
\email{leeyating@hust.edu.cn}

\author{Ruqian Zhang}
\authornotemark[1]
\affiliation{%
 \institution{School of Cyber Science and Engineering, Huazhong University of Science and Technology}
 \city{Wuhan}
 \country{China}
 }
\email{ruqianzhang@hust.edu.cn}

\author{Shouhuai Xu}
\affiliation{%
 \institution{Department of Computer Science, University of Colorado Colorado Springs, Colorado Springs}
 \country{Colorado, USA}
 }
\email{sxu@uccs.edu}

\author{Chao Zhang}
\authornotemark[3]
\affiliation{%
 \institution{Institute for Network Sciences and Cyberspace, Tsinghua University}
 \city{Beijing}
 \country{China}
 }
\email{chaoz@tsinghua.edu.cn}

\author{Hai Jin}
\authornotemark[1]
\authornotemark[2]
\affiliation{%
 \institution{School of Computer Science and Technology, Huazhong University of Science and Technology}
 \city{Wuhan}
 \country{China}
 }
\email{hjin@hust.edu.cn}

\renewcommand{\shortauthors}{Zhen Li, Ning Wang, Deqing Zou, Yating Li, Ruqian Zhang, Shouhuai Xu, Chao Zhang, and Hai Jin}

\begin{abstract}
Software vulnerabilities are a major cyber threat and it is important to detect them. 
One important approach to detecting vulnerabilities is to use deep learning while treating a program function as a whole, known as {\em function-level} vulnerability detectors.
However, the limitation of this approach is not understood. In this paper, we investigate its limitation in detecting one class of vulnerabilities known as
{\em inter-procedural vulnerabilities}, where the {\em to-be-patched statements} and the {\em vulnerability-triggering statements} belong to different functions. 
For this purpose, we create the first {\em {\underline {Inter}-{\underline P}rocedural {\underline V}ulnerability {\underline D}ataset}} (InterPVD) based on C/C++ open-source software, and
we propose a tool dubbed VulTrigger for identifying vulnerability-triggering statements across functions. 
Experimental results show that VulTrigger 
can effectively identify vulnerability-triggering statements and inter-procedural vulnerabilities.
Our findings include: 
(i) inter-procedural vulnerabilities are prevalent with an average of 2.8 inter-procedural layers;  and
(ii) function-level vulnerability detectors are much less effective in 
detecting to-be-patched functions of 
inter-procedural vulnerabilities than detecting their counterparts of  
intra-procedural vulnerabilities.
\end{abstract}




\maketitle

\section{Introduction}
Software vulnerabilities are arguably the most significant 
cyber threats.
For instance, the Apache Log4j2 vulnerability (CVE-2021-44228) can be exploited to launch 
a remote code execution attack, which has had a huge impact because Apache Log4j2
is widely employed by many enterprises.
This highlights the importance of detecting software vulnerabilities. 
One widely-used approach to detecting software vulnerabilities in source code is to use  {\em Static Application Security Testing} (SAST) tools, including open-source tools \cite{Flawfinder,CodeQL,Infer}
and commercial tools \cite{Checkmarx,Fortify}.
However, the false-positive and false-negative rates of these SAST tools are high because of their incomplete vulnerability rules or patterns 
defined by human experts \cite{DBLP:conf/issta/LippBP22,VulChecker}. To avoid these weaknesses, researchers have investigated deep learning-based vulnerability detectors.

One promising approach is to encode a program function as a whole, known as
{\em function-level} vulnerability detectors
\cite{DBLP:conf/ijcai/DuanWJRLYW19,DBLP:conf/nips/ZhouLSD019,DBLP:journals/tifs/WangYTTHFFBW21,DBLP:conf/sigsoft/Li0N21,DBLP:conf/icse/WuZD0X022,DBLP:conf/ijcnn/HanifM22,DBLP:conf/msr/FuT22,DBLP:journals/tse/ChakrabortyKDR22}.
This approach can effectively detect {\em intra-procedural vulnerabilities}, where both the {\em to-be-patched statements} (that must be patched to eliminate a vulnerability) and the {\em vulnerability-trigger-ing statements}\footnote{
We do not use the general term of ``vulnerable statements'' because some researchers treat both {\em to-be-patched statements} and {\em vulnerability-triggering statements} as vulnerable statements. 
It is important to 
distinguish between these statements in this paper.} 
(that trigger the vulnerability but do not need to be patched) belong to the same function.
However, the limitation of this approach to detecting another class of vulnerabilities known as {\em inter-procedural vulnerabilities}, where the to-be-patched statements and the vulnerability-triggering statements belong to {\em different} functions, is not understood. It is important to characterize this limitation because their incapability in detecting the vulnerability-triggering statements may cause false-negatives and/or make it hard to explain why the to-be-patched statements (even if detected) represent a vulnerability. 
In the latter case, practitioners may simply ignore these detected
vulnerabilities 
\cite{DBLP:journals/tosem/ZouZXLJY21,DBLP:conf/sigsoft/Li0N21,DBLP:conf/ccs/GuoMXSWX18}. 

\smallskip
\noindent{\bf State-of-the-art: Lack of datasets and effective detectors for inter-procedural vulnerabilities.} 
On one hand, although existing vulnerability datasets \cite{DBLP:conf/dsn/WangWF0J21,DBLP:conf/msr/FanL0N20,DBLP:conf/promise/BhandariNM21,DBLP:conf/sigsoft/NikitopoulosDLM21}
do provide vulnerability patches from which to-be-patched statements can be obtained, they do not provide any vulnerability-triggering statements. This explains why
studies leveraging such datasets \cite{DBLP:conf/ccs/LiP17,DBLP:conf/icse/LiuMZGLLSH020,DBLP:conf/cns/WangW0BJ20} cannot identify inter-procedural vulnerabilities, despite that they can answer the question whether or not multiple functions are involved in a vulnerability patch.
This also suggests that these studies could mistakenly claim that a vulnerability involves only one vulnerable function.

On the other hand, we are not aware of any detectors that can effectively identify inter-procedural vulnerabilities.
This is true despite that some SAST tools \cite{Flawfinder,Infer,CodeQL,Checkmarx,Fortify} can indeed identify some vulnerability-triggering statements, but their accuracy is too low to be useful
because they would miss 47\%-80\% of vulnerabilities \cite{DBLP:conf/issta/LippBP22} and can achieve at most a 28.8\% accuracy in identifying vulnerability-triggering statements (as shown by our study which will be presented in Table \ref{Table_accuracy_sink} and Table \ref{Table_compiling_tools} in Section \ref{subsec:RQ1}).
Dynamic analysis methods (e.g., fuzzing \cite{DBLP:journals/tse/ManesHHCESW21,DBLP:conf/sp/GanZQTLPC18,DBLP:conf/icse/MengDLBR22}) have high false-negative rates because they cannot test all execution paths despite that they can identify vulnerability-triggering statements.
It is also worth mentioning that dynamic analysis tools incur a very high overhead to trigger a vulnerability, meaning that they cannot scale up to deal with a large number of programs. 

\noindent{\bf Our contributions.} In this paper, we present the first study on the effectiveness 
of function-level vulnerability detectors in detecting inter-procedural vulnerabilities.
Specifically, we make three contributions.
First, we propose 
a novel method and accompanying tool, dubbed VulTrigger, to automatically identify vulnerability-triggering statements for known vulnerabilities with patches. 
VulTrigger characterizes 10 popular vulnerability types (involving 16 CWEs) and can identify 19 types of vulnerability-triggering statements by leveraging (i) vulnerability patches to obtain {\em critical variables} and (ii) a newly proposed program slicing method, which may be of independent value.
Experimental results show that VulTrigger significantly outperforms the SAST tools \cite{Flawfinder,Checkmarx,Infer,CodeQL,Fortify}, which are the existing tools that have some capabilities in identifying vulnerability-triggering statements at scale, 
by achieving a 55.8\% higher accuracy on average when applied to identify vulnerability-triggering statements. VulTrigger also outperforms the other existing methods \cite{DBLP:conf/ccs/LiP17,DBLP:conf/icse/LiuMZGLLSH020,DBLP:conf/cns/WangW0BJ20}, which have some capabilities in identifying inter-procedural vulnerabilities, with an improvement of 20.6\% in 
the false positive rate, 32.7\% in the false negative rate, and 45.1\% in the overall effectiveness F1-measure when applied to identify inter-procedural vulnerabilities.

Second, we apply VulTrigger to build the first {\em {\underline {Inter}}-{\underline P}rocedural {\underline V}ulnerability {\underline D}ataset} (InterPVD) based on C/C++ open-source software, which may be of independent value. 
The dataset involves 769 vulnerabilities, each of which is labeled with its patch statements, its vulnerability-triggering statements, its nature in terms of being inter-procedural or not, its type of inter-procedural vulnerability, and its sequence of functions starting from the to-be-patched function to the vulnerability-triggering function. 
We find that 24.3\% of the vulnerabilities in the InterPVD are inter-procedural vulnerabilities with 2.8 layers on average, 
meaning that inter-procedural vulnerabilities are prevalent in open-source software.

Third, we investigate the effectiveness of 5 state-of-the-art function-level vulnerability detectors via InterPVD. 
Experimental results show: 
(i) function-level vulnerability detectors are much less effective in detecting vulnerability-triggering functions than detecting to-be-patched functions; 
(ii) detecting to-be-patched functions of inter-procedural vulnerabilities is more challenging than detecting their counterparts of intra-procedural vulnerabilities.

We have published our dataset InterPVD, VulTrigger source code, and other tools we evaluated at \url{https://github.com/CGCL-codes/VulTrigger}.

\section{Inter-Procedural Vulnerabilities}
\label{sec:characterization}
\input{Characterization.tex}

\section{VulTrigger}
\label{sec:methodology}
\input{VulTrigger.tex}

\section{Generating and Analyzing InterPVD}
\label{sec:dataset}
\input{InterPVD.tex}

\section{Effectiveness of Function-Level Vulnerabilities Detectors 
}
\label{sec:evaluation}

\input{Evaluation.tex}

\section{Discussion}
\label{sec:discussion}
\input{Discussion.tex}

\section{Related Work}
\label{sec:related_work}
\input{RelatedWork.tex}

\section{Conclusion}
\label{sec:conclusion}
We have investigated the effectiveness 
of function-level vulnerability detectors in coping with inter-procedural vulnerabilities in C/C++ open-source software, by proposing the innovative VulTrigger to identify vulnerability-triggering statements for known vulnerabilities with patches.
We show that inter-procedural vulnerabilities are prevalent in C/C++ open-source software, and that detecting to-be-patched functions and vulnerability-triggering functions, especially the latter, for inter-procedural vulnerabilities is significantly more challenging than for intra-procedural vulnerabilities. This explains why existing function-level vulnerability detectors cannot effectively cope with inter-procedural vulnerabilities. 
The limitations discussed in Section \ref{sec:discussion} offer interesting problems for future research.

\section{Acknowledgement}
\label{sec:acknowledgement}
We thank the anonymous reviewers for their comments which guided us in improving the paper. The authors affiliated with Huazhong University of Science and Technology were supported by the National Natural Science Foundation of China under Grant No. 62272187. Shouhuai Xu was partly supported by the National Science Foundation under Grants \#2122631, \#2115134, and \#1910488 as well as Colorado State Bill 18-086. Any opinions, findings, conclusions, or recommendations expressed in this work are those of the authors and do not reflect the views of the funding agencies in any sense.

\input{bibliography.bbl}

\bibliographystyle{ACM-Reference-Format}
\bibliography{bibliography}

\end{document}

%% file: Characterization.tex
\subsection{Vulnerability-Triggering Statements}
Figure \ref{Fig_cross_function_CVE_example}(a) describes an example of inter-procedural vulnerability corresponding to CVE-2015-8662. In this example, {\tt ff\_dwt\_decode} is the {\em to-be-patched function} (i.e., the function containing some to-be-patched statement(s)); the {\em out-of-bounds array access} vulnerability is triggered by the {\em vulnerability-triggering statement} in the {\em vulnerability-triggering function}, 
which is the Line 329 in another function, namely {\tt dwt\_decode53}. For this example, the function-level vulnerability detectors may treat {\tt ff\_dwt\_decode} alone 
as vulnerable. This contrasts the fact that whether {\tt ff\_dwt\_decode} is vulnerable or not depends on the presence of Line 329 in {\tt dwt\_decode53}. 
Figure \ref{Fig_cross_function_CVE_example}(b) describes the diff file for patching the vulnerability,
where the three lines of code in green color are called {\em patch statements}. In this case, the patch is to add
control statements.
In general, patch statements can be defined as:

\begin{figure}[!t]
	\vspace{-0.2cm}
	\centering
	\includegraphics[width=0.34\textwidth]{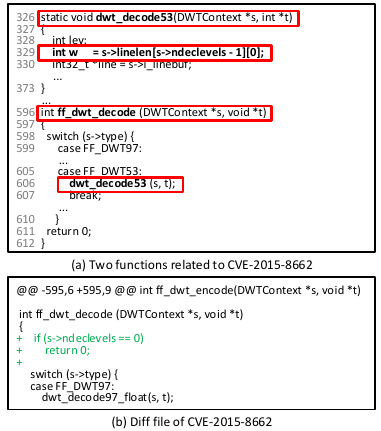}
	\vspace{-0.3cm}
	\caption{An example of inter-procedural vulnerability (CVE-2015-8662) with to-be-patched function {\tt ff\_dwt\_decode}, vulnerability-triggering function {\tt dwt\_decode53}, and vulnerability-triggering statement (Line 329) 
 }
	\vspace{-0.2cm}
	\label{Fig_cross_function_CVE_example}
\end{figure}

\vspace{-0.15cm}
\begin{definition}[Patch statements]
\label{definition:patch}
Given a vulnerability $v$ 
and its patch via diff file $d$, its {\em patch statements} are the set of statements $P$ that are added, deleted, or modified 
as per $d$. 
\end{definition}
\vspace{-0.15cm}

According to Definition \ref{definition:patch}, a {\em to-be-patched function} is a function that must be patched by incorporating one or multiple patch statements. 
If the patch statement(s) belong to
multiple functions, these functions are also to-be-patched functions.
Patches can correspond to adding, deleting, and modifying statements, which may involve variable assignments, system calls, variable definitions, function definitions, and control statements. This leads to
the 11 types of patch statements, dubbed P-1 to P-11 and highlighted in Table \ref{Table_patch_patterns}.

\begin{table}[!t]
	\caption{Types of patch statements
	}
	\vspace{-0.4cm}
	\label{Table_patch_patterns}
	\footnotesize
	\centering
	\begin{tabular}{|c|l|}
		\hline
		No. & Description  \\
		\hline
		P-1 & Add/delete variable assignments\\
		\hline
		P-2 & Modify variable assignments \\
		\hline
		P-3 & Add/delete function calls \\
		\hline
		P-4 & Modify function calls\\
		\hline
		P-5 & Add/delete variable definitions \\
		\hline
		P-6 & Modify variable definitions \\
		\hline
		P-7 & Add/delete function definitions \\
		\hline
		P-8 & Modify function definitions \\
		\hline
		P-9 & Add/delete control statements \\
		\hline
		P-10 & Modify control statements \\
		\hline
		P-11 & Others \\
		\hline
	\end{tabular}
	\vspace{-0.4cm}
\end{table}

\begin{definition}[Vulnerability-triggering statements]
\label{definition:sink}
Given a vulnerability $v$, the vulnerability-triggering statements of $v$ are a set of statements $T$ such that $v$ is triggered for the {\em first} time on an execution path; that is, an incorrect program state is manifested as a consequence of executing the vulnerability-triggering statement. 
\end{definition}

According to Definition \ref{definition:sink}, a {\em vulnerability-triggering function} is a function that contains one or multiple vulnerability-triggering statements.
Note that Definition \ref{definition:sink} only considers the {\em first} time when a vulnerability is triggered on an execution path because the program behaves abnormally after exploitation.

We focus on 10 common vulnerability types (involving 16 CWEs \cite{CWE}) listed in Table \ref{Table_cwe_sink_type}, because these vulnerability types can provide more vulnerabilities with diff files which are collected from the {\em National Vulnerability Database} (NVD) \cite{NVD}.
By considering the scenario that multiple vulnerabilities belong to the same CWE but may be triggered in different fashions and the scenario that multiple vulnerabilities belong to different CWEs but may be triggered in the same fashion, we define 19 types of vulnerability-triggering statements for the above-mentioned 16 CWEs according to the vulnerability-triggering fashion. 
Table \ref{Table_cwe_sink_type} shows the correspondence between 16 CWEs corresponding to the 10 vulnerability types and the 19 types of vulnerability-triggering statements described in Table \ref{Table_sink_type}. Note that a vulnerability-triggering statement may belong to multiple types of vulnerability-triggering statements. For example, a macro definition related to division (T-19) can be also a division-by-zero statement (T-17).

Some types of vulnerabilities have no obvious vulnerability-triggering statements, in which case we select the statement(s) that are mostly related to the vulnerability as the vulnerability-triggering statements. 
Take vulnerabilities of the {\em missing release of resources} type for example, we define the last executed statements in the to-be-patched function as the vulnerability-triggering statements, which often involve the {\tt return} statement or the last statement of the to-be-patched function in the absence of the {\tt return} statement. This is reasonable because the resources should be released before the to-be-patched function ends.
In what follows, we elaborate on two examples of vulnerability-triggering statements.

\begin{table}[!tb]
	\vspace{-0.2cm}
	\caption{Vulnerability types and their corresponding types of vulnerability-triggering statements}
	\vspace{-0.4cm}
	\label{Table_cwe_sink_type}
	\scriptsize
	\centering
	\begin{tabular}{|c|c|m{.13\textwidth}<{\centering}|}
		\hline
		Vulnerability type & CWE ID & Types of vulnerability-triggering statements \\
		\hline
		{\multirow{4}{*}{Buffer overflow}} & CWE-119 & T-1, T-2, T-3 \\
		\cline{2-3}
		& CWE-125 & T-1, T-2, T-3 \\
		\cline{2-3}
		& CWE-787 & T-1, T-2, T-3 \\
		\cline{2-3}
		& CWE-120 & T-1, T-2, T-3 \\
		\hline
		{\multirow{3}{*}{Numeric error}} & CWE-189 & T-1, T-2, T-3, T-4 \\
		\cline{2-3}
		& CWE-190 & T-1, T-2, T-3, T-4 \\
		\cline{2-3}
		& CWE-191 & T-1, T-2, T-3, T-4 \\
		\hline
		Reachable assertion & CWE-617 & T-5\\
		\hline
		Path traversal & CWE-22 & T-6\\
		\hline
		Infinite loop & CWE-835 & T-7, T-8, T-9\\
		\hline
		{\multirow{2}{*}{Missing release of resources}} & CWE-772 & T-10, T-11\\
		\cline{2-3}
		& CWE-401 & T-10, T-11\\
		\hline
		Double-free & CWE-415 & T-12 \\
		\hline
		Use-after-free & CWE-416 & T-12, T-13 \\
		\hline
		NULL pointer dereference & CWE-476 & T-14, T-15, T-16\\
		\hline
		Division-by-zero & CWE-369 & T-17, T-18, T-19\\
		\hline
	\end{tabular}
	\vspace{-0.2cm}
\end{table}

\begin{table}[!tb]
	\caption{Types of vulnerability-triggering statements
	}
	\vspace{-0.4cm}
	\label{Table_sink_type}
	\scriptsize
	\centering
	\begin{tabular}{|c|m{.41\textwidth}|}
		\hline
		No. & Description \\
		\hline
		T-1 & API or system call-related function call (e.g., {\tt memcpy}, {\tt alloc}, {\tt memset}) that may trigger out-of-bounds access or memory errors \\
		\hline
		T-2 & Out-of-bounds array access  \\
		\hline
		T-3 & Out-of-bounds pointer access \\
		\hline
		T-4 & Integer overflow or underflow caused by integer addition, subtraction and multiplication \\
		\hline
		T-5 & Assertion function calls (e.g., {\tt assert} and {\tt BUG}) \\
		\hline
		T-6 & Path access-related API/system calls (e.g., {\tt open}, {\tt read}, {\tt path\_copy}, and {\tt mkdir})\\
		\hline
		T-7 & Loop condition of {\tt for}, {\tt while}, and {\tt do while} \\
		\hline
		T-8 & {\tt goto} statement \\
		\hline
		T-9 & Recursive function call\\
		\hline
		T-10 & {\tt return} statement \\
		\hline
		T-11 & The last statement of the to-be-patched function in the absence of the {\tt return} statement\\
		\hline
		T-12 & Function calls related to {\tt free}, {\tt delete}, {\tt destroy}, and {\tt unregister} \\
		\hline
		T-13 & The first usage of a critical variable when there are no function calls related to {\tt free}, {\tt delete}, {\tt destroy}, and {\tt unregister} \\
		\hline
		T-14 & The first usage of a member of a critical variable, which is a struct type without initialization \\
		\hline
		T-15 & API or system call-related memory allocation function calls (e.g., {\tt memcpy} and {\tt alloc}) when the critical variable is not a variable of  struct type and its value is NULL or not assigned \\
		\hline
		T-16 & The uninitialized function is called, mainly in the case of function pointers \\
		\hline
		T-17 & Division-by-zero such as $/c$ and $\%c$ where $c$ is a critical variable \\
		\hline
		T-18 & API or system call related to {\tt alloc} \\
		\hline
		T-19 & The usage of macro definition related to division \\
		\hline
	\end{tabular}
	\vspace{-0.4cm}
\end{table}

\begin{figure}[!tb]
	\centering
	\includegraphics[width=.35
	\textwidth]{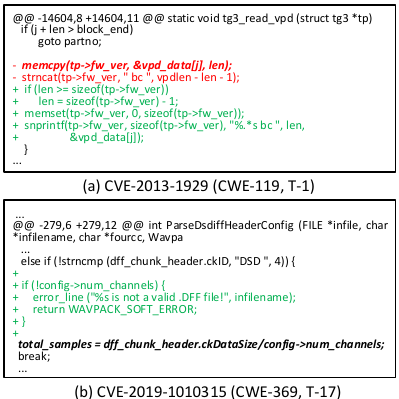}
	\vspace{-0.3cm}
\caption{Diff files of vulnerabilities with 
vulnerability-triggering statements (highlighted in bold and italics)
}
	\vspace{-0.4cm}
	\label{Fig_sink_cve_example}
\end{figure}

\noindent{\bf Type 1 (T-1): API or system call-related function call.}
For buffer-overflow vulnerabilities, the vulnerability-triggering statement of T-1 is the API or system call-related function call that may trigger out-of-bounds access or memory error, such as {\tt memcpy}, {\tt alloc}, or {\tt memset}. Consider CVE-2013-1929 (CWE-119) in Figure \ref{Fig_sink_cve_example}(a) as an example. The variable {\tt len} exceeds the appropriate length when copying {\tt vpd\_data[j]} to {\tt tp->fw\_ver} in function {\tt memcpy}, which causes an out-of-bounds access.
The vulnerability-triggering statement is ``{\tt memcpy(tp->fw\_ver, \&vpd\_data[j], len);}''.

\noindent{\bf Type 17 (T-17): division-by-zero.} For division-by-zero vulnerabilities, the vulnerability-triggering statement of T-17 is division-by-zero, such as $/c$ and $\%c$ where $c$ is a critical variable (i.e., a variable related to the vulnerability). Take the vulnerability CVE-2019-1010315 (CWE-369) in Figure \ref{Fig_sink_cve_example}(b) as an example. The value of {\tt config->num\_channels} can be zero,
causing a division-by-zero error. The vulnerability-triggering statement is ``{\tt total\_samples=diff\\\_chunk\_header.ckDataSize/config->num\_channels;}''.

\vspace{-0.2cm}
\subsection{Characterizing Vulnerability-Triggering Statements}
\label{subsec:Step3}
To obtain characteristics, we analyze the vulnerability-triggering statements corresponding to the 16 CWEs described in Table \ref{Table_cwe_sink_type}. 
We divide the vulnerability types into {\em two classes} according to whether the vulnerability-triggering statements are closely related to the critical variables. 
The {\em first class} of vulnerability types corresponds to the vulnerability-triggering statements that are not closely related to critical variables but do involve {\em missing release of resource} (CWE-772 and CWE-401) or {\em infinite loop} (CWE-835). Take characteristics related to missing release of resource as an example.
In a to-be-patched function, if there exists a {\tt return} statement in either the patch statements or their subsequent statements, the vulnerability-triggering statement is the {\tt return} statement, which is one of the patch statements or the statement closest to the patch statements (T-10). The {\tt return} statement becomes the last statement executed in the function when the vulnerability is triggered. Otherwise, if there is no {\tt return} statement, the vulnerability-triggering statement is the last statement in the to-be-patched function (T-11).

The {\em second class} of vulnerability types corresponds to the vulnera-
bility-triggering statements that depend on the critical variables, involving buffer overflow, numeric error, reachable assertion, path traversal, double free, use after free, NULL pointer dereference, and division-by-zero. Take characteristics related to a buffer overflow as an example. 
We create a list of keywords of APIs/system calls, which may trigger out-of-bounds access or memory errors. The first type of vulnerability-triggering statements is the ones in the program slice that involve both the critical variable and the function call containing a keyword in the list of keywords (T-1). If the critical variable is used as an array name or an array index, the vulnerability-triggering statement is the usage of the array involving the critical variable in the program slice (T-2). If the critical variable is a pointer, the vulnerability-triggering statement is the statement in the program slice that involves the pointer that accesses memory or is involved in arithmetic operations (T-3).

\vspace{-0.1cm}
\subsection{Inter-Procedural Vulnerabilities}
Based on patch and vulnerability-triggering statements, we define the inter-procedural vulnerability as follows.

\vspace{-0.1cm}
\begin{definition}[Inter-procedural vulnerability]
\label{definition:cross_func}
Consider a vulnerability $v$, its patch statements $P$, and vulnerability-triggering statements $T$. If a patch statement $p \in P$ belongs to function $f$, a vulnerability-triggering statement $t \in T$ belongs to another function $g$ ($f \neq g$), and $t$ is data- or control-dependent on $p$, then we say $v$ is an {\em inter-procedural vulnerability}. The number of inter-procedural layers is the number of different functions 
involved in the data-flow or control-flow from  $p$ to $t$.
\end{definition}
\vspace{-0.1cm}

It is worth mentioning that for the vulnerability whose patch statements exist in multiple to-be-patched functions 
(corresponding to one or multiple vulnerability-triggering functions), 
if there is a to-be-patched function which is different from its corresponding vulnerability-triggering function, the vulnerability is an inter-procedural vulnerability. 

\begin{figure}[!tb]
	\centering
	\includegraphics[width=.42\textwidth]{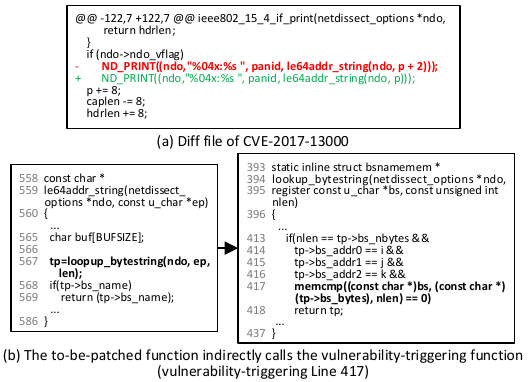}
	\vspace{-0.3cm}
	\caption{An inter-procedural vulnerability  of {\em caller type} corresponding to CVE-2017-13000 (CWE-125)}
	\vspace{-0.2cm}
	\label{Fig_cross_function_type_1}
\end{figure}

According to the relationship between the to-be-patched function $f$ and the vulnerability-triggering function $g$, there are two types of inter-procedural vulnerabilities.  
The first type 
is that the to-be-patched function $f$ directly or indirectly calls the vulnerability-triggering function $g$, dubbed ``caller type''. 
Take the vulnerability CVE-2017-13000 (CWE-125) in Figure \ref{Fig_cross_function_type_1} as an example. The to-be-patched function {\tt ieee802\_15\_4\_if\_print} calls function {\tt le64addr\\\_string}, which calls function {\tt loopup\_bytestring} at Line 567. The vulnerability-triggering statement {\tt memcmp((const char *)bs, (const char *)(tp->bs\_bytes), nlen) == 0} (Line 417) in function {\tt loopup\_bytestring} may cause out-of-bounds read when reading {\tt nlen} bytes from {\tt bs} and {\tt tp->bs\_byte}.
The number of inter-procedural layers is 3, corresponding to three functions {\tt ieee802\_15 \_4\_if\_print}, {\tt le64addr\_string}, and {\tt loop-up\_bytestring}.

\begin{figure}[!tb]
	\vspace{-0.2cm}
	\centering
	\includegraphics[width=.42\textwidth]{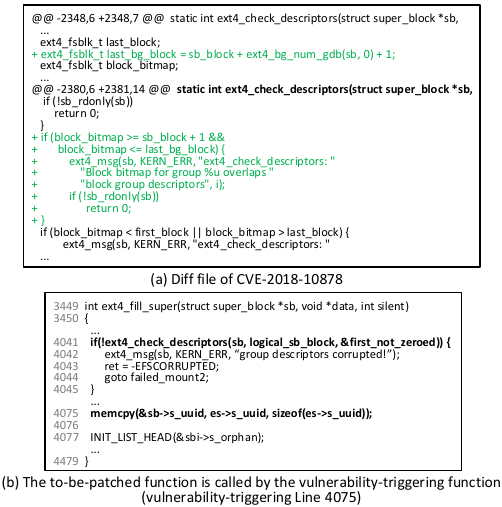}
	\vspace{-0.3cm}
	\caption{An inter-procedural vulnerability of {\em callee type} corresponding to CVE-2018-10878 (CWE-787)}
	\vspace{-0.4cm}
	\label{Fig_cross_function_type_2}
\end{figure}

The second type of inter-procedural vulnerabilities is that 
the vulnerability-triggering function $g$ may appear anywhere after the to-be-patched function $f$ returns, dubbed ``callee type''.
Take the vulnerability CVE-2018-10878 (CWE-787) in Figure \ref{Fig_cross_function_type_2} as an example. Function {\tt ext4\_fill\_super} calls to-be-patched function {\tt ext4\_check\_descrip}- {\tt tors}, then triggers the vulnerability after the to-be-patched function returns, causing out-of-bounds write at Line 4,075.
The number of inter-procedural layers is 2, corresponding to 
functions {\tt ext4\_fill\_super} and {\tt ext4\_check\_descriptors}.

%% file: VulTrigger.tex
\begin{figure}[!tb]
	\vspace{-0.2cm}
	\centering
	\includegraphics[width=0.38\textwidth]{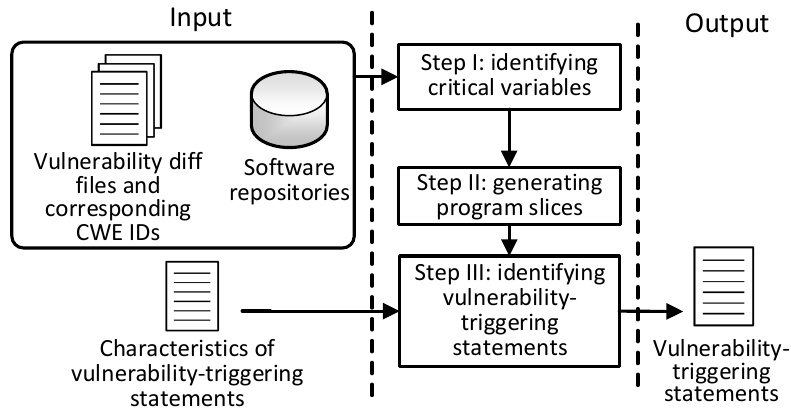}
	\vspace{-0.3cm}
	\caption{Overview of VulTrigger}
	\vspace{-0.6cm}
	\label{Fig_overview}
\end{figure}

As highlighted in Figure \ref{Fig_overview}, we propose VulTrigger for identifying vulnerability-triggering statements of vulnerabilities. 
At a high level, VulTrigger takes as input some vulnerability diff files, their corresponding CWE IDs, software repositories, and characteristics of vulnerability-triggering statements; it identifies critical variables based on the patch statements (Step I); then, it generates program slices corresponding to these critical variables (Step II); after that, it identifies the vulnerability-triggering statements (Step III); finally, it outputs vulnerability-triggering statements.

\begin{figure*}[!t]
	\vspace{-0.2cm}
	\centering
	\includegraphics[width=\textwidth]{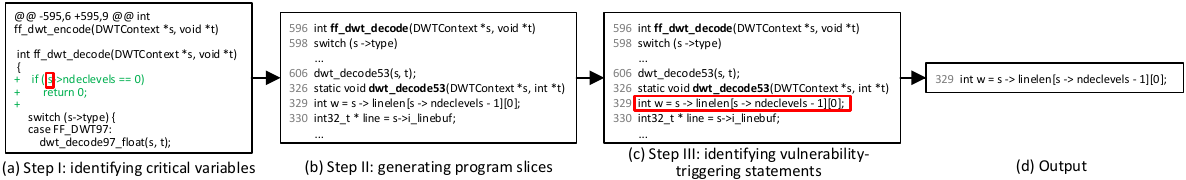}
	\vspace{-0.7cm}
	\caption{Using CVE-2015-8662 in Figure \ref{Fig_cross_function_CVE_example} to illustrate the identification of vulnerability-triggering statements,
	where the two red boxes respectively highlight the identified critical variable and the vulnerability-triggering statement.}
	\vspace{-0.4cm}
	\label{Fig_program_slicing_example}
\end{figure*}

\subsection{Identifying Critical Variables 
(Step I)}
The purpose of this step is to identify critical variables based on patch statements. Given a vulnerability and its diff file, the starting point for tracing how the vulnerability can be triggered is to identify the {\em critical variables} related to the vulnerability. 
We first preprocess the diff file to remove the comments, the blank lines of code, and some common semantically equivalent statement modifications in the patch statements. 
Then, we identify critical variables for each patch statement according to the types of patch statements as described in Table \ref{Table_patch_patterns}. 

\noindent{\bf Variable assignments.} For P-1 
and P-2 patch statements, 
critical variables are usually the variables to which values are assigned. For numeric error vulnerabilities, 
we treat all variables in the assignment statements as critical variables because numeric error vulnerabilities can be triggered at the assigned variables on the left-hand side of the assignment symbol or the arithmetic operations of variables on the right-hand side of the assignment symbol.

\noindent{\bf Function calls.} For P-3 patch statements, 
the critical variables are the argument variables
of the function call. 
For P-4 patch statements, 
the critical variables are the modified argument variables of the function call.

\noindent{\bf Variable definitions.} For  P-5 patch statements, 
the critical variables are the added/deleted variables.
For P-6 patch statements, 
the critical variables are the variables whose definitions are modified.

\noindent{\bf Function definitions.} For P-7 patch statements, 
the critical variables are the argument variables of those functions whose definitions are added/deleted. 
For P-8 patch statements, 
the critical variables are the modified argument variables of these functions. 

\noindent{\bf Control statements.} For P-9 patch statements, 
the critical variables are the variables in the control conditions.
For P-10 patch statements, 
the critical variables are the modified variables in the control statements.

\noindent{\bf Others.} For other types of patch statements, we take {\tt return} statements as an example. 
If the patch statements only involve a {\tt return} statement addition/deletion, the critical variables are the variables that are involved in the {\tt return} statements; otherwise, the critical variables are the modified variables in the {\tt return} statements.

\noindent{\bf Running example}. In any case, if a critical variable is a member of struct type, the struct type variable is considered a critical variable. For the vulnerability (CVE-2015-8662) shown in Figure \ref{Fig_cross_function_CVE_example}, Figure \ref{Fig_program_slicing_example}(a) shows the critical variable identified in Step I. Since the patch statements only involve adding control statements, the type of patch statements is P-9. The critical variable is 
a struct type variable {\tt s}.

\subsection{Generating Program Slices 
(Step II)}
The purpose of this step is to generate program slices corresponding to the critical variables identified in Step I. To accomplish this, we first extract direct and indirect dependency files of to-be-patched functions (via the system dependency graph of software), which may call the other functions or use the variables or types that are defined in other program files. 
Then, we generate inter-procedural program slices corresponding to the critical variables as follows.
(i) In the case the patch statements in the preprocessed diff file are only about adding statements, we generate program slices for the patched version of the to-be-patched function and the functions that directly or indirectly call the to-be-patched function and that are directly or indirectly called by the to-be-patched function in the patched program. We further delete from each program slice the statements that do not appear in the vulnerable program.
(ii) In the case the patch statements in the preprocessed diff file are about deleting or modifying statements, we generate program slices for the to-be-patched function 
and the functions that directly or indirectly call the to-be-patched function and that are directly or indirectly called by the to-be-patched function in the vulnerable program.
Our method for generating program slices incorporates the following three improvements based on the method presented in \cite{li2021sysevr}.

\noindent{\bf Improvement in dealing with control statements.}
It is known that when the patch statement is a control statement, such as {\tt if}, {\tt for}, or {\tt while}, forward slicing according to the control dependency may result in too many statements while forward slicing according to the data dependency may miss some vulnerabilities
\cite{li2021sysevr}. 
Our improvement lies in the following steps. 
We first obtain the critical variables in the control statement, then conduct the backward slicing till 
the definition of critical variables, and further conduct forward slicing according to the 
critical variables with respect to data dependency. 
Finally, we remove the statements between the statement of the critical variable definition
and the control statement from the program slice to exclude the statements that are irrelevant to the vulnerability.
This improvement assures that the data-dependent statements related to the critical variable (thus the vulnerability) can be captured in the program slice without introducing irrelated statements.

\noindent{\bf Improvement in dealing with implicit return values of function calls.}
For method \cite{li2021sysevr}, if a function call does not explicitly return values or returns an error, the resulting program slice will end with the function call statement. This is problematic because a pointer variable, when used as an argument in a function call, is an implicit return value, which should have data dependency with the following usage of the pointer variable. To resolve this problem, we identify the pointer variable arguments and conduct backward slicing till
the definition of the pointer variable. Then, we conduct forward slicing according to the pointer variable with respect to data dependency. Finally, we remove the statements between the statement of the pointer variable definition and the function call statement from the program slice to exclude statements that are irrelevant to the vulnerability.

\noindent{\bf Improvement in dealing with program slicing after the to-be-patched function returns.} The program slicing method \cite{li2021sysevr} can work with the caller type of inter-procedural vulnerability. 
However, the 
method \cite{li2021sysevr} stops when the to-be-patched function ends. 
In this paper, we need to obtain the program slices after the to-be-patched function returns. To achieve this, we generate the function call graph of the vulnerable program to obtain a set of functions, denoted by $H$, which call the to-be-patched function $f$. For each function $h \in H$, we conduct the inter-procedural program slicing as in \cite{li2021sysevr} by starting at the to-be-patched function call statement. Going beyond the method \cite{li2021sysevr}, 
we let program slicing continue in function $h$ after the to-be-patched function returns. To obtain the statements of program slices after function $h$ returns, the process is similar to that of obtaining the program slices after function $f$ returns.
This procedure is iterated until the number of inter-procedural layers reaches a given threshold $\theta$.

\noindent{\bf Running example}.
Corresponding to CVE-2015-8662, 
Figure \ref{Fig_program_slicing_example}(b) depicts a program slice corresponding to critical variable 
{\tt s}. The patch statements include  control statement {\tt if(s->ndeclevels == 0)}. 
{\color{black}The program slice is generated by initially extracting statements from the patched version, and then deleting the statements that do not appear in the vulnerable program.}
The process of program slicing involves a function call without explicit return values, namely ``{\tt dwt\_decode53 (s, t);}''.
The generated program slice corresponding to the vulnerability includes two functions: {\tt ff\_dwt\_decode} and {\tt dwt\_decode53}.

\subsection{Identifying Vulnerability-Triggering Statements (Step III)}
Given a vulnerability and its CWE ID, we 
identify vulnerability-triggering statements for each program slice generated in Step II.
For the first class of vulnerability types (CWE-835, CWE-772, and CWE-401), we identify vulnerability-triggering statements by leveraging characteristics of vulnerability-triggering statements of the CWE ID. 
For the second class of vulnerability types, there are three substeps. 
(i) For the critical variables obtained from Step I (denoted by ``the first-level critical variables''), we identify the vulnerability-triggering statements from the patch statements or the subsequent statements in the program slice according to the characteristics of the vulnerability-triggering statements corresponding to the CWE ID. 
(ii) If there are no vulnerability-triggering statements identified in the previous step, we transform the first-level critical variables to other related critical variables as the second-level critical variables. The transformation strategies are: 
\begin{itemize}
[leftmargin=.32cm,noitemsep,topsep=2pt]
\item {\bf Assignment.} 
If a critical variable $c$ is assigned to another variable $c'$, then $c'$ is the second-level critical variable. 
\item {\bf Function header in the function definition.} 
If the statement is a function header in the program slice, we trace back to the place where the function is called. If the critical variable is an argument of the function call, we transform the critical variable to the corresponding parameter variable in the function definition.

\item {\bf To-be-patched function call with return value.} 
If the return value of the to-be-patched function call involves the value of a critical variable, we transform the critical variable to the variable that is assigned with the return value.
\end{itemize}

Then, we transform second-level critical variables to third-level critical variables, and so on. The smaller the level a variable has, the higher the priority. The larger-level variables are used to identify vulnerability-triggering statements 
only when the smaller-level critical variables fail to identify vulnerability-triggering statements.
(iii) If no vulnerability-triggering statements are identified in the previous two steps, we identify vulnerability-triggering statements from 
the statements preceding the patch statements in the program slice according to the characteristics of vulnerability-triggering statements. This situation is assured to be encountered 
because of loop structures and recursive function calls.

\noindent{\bf Running example.}
Corresponding to 
CVE-2015-8662 shown in Figure \ref{Fig_cross_function_CVE_example}, the vulnerability type is CWE-119. A member of the critical variable {\tt s} is used as the index of an array {\tt s->linelen}, and the vulnerability-triggering statement is the usage of the array involving the critical variable in the program slice (Line 329, T-2).
Figure \ref{Fig_program_slicing_example}(c) shows the identified vulnerability-triggering statement.

%% file: InterPVD.tex
\subsection{Applying VulTrigger to Construct InterPVD}
We focus on 16 CWEs described in Table \ref{Table_cwe_sink_type} and collect the open-source C/C++ vulnerabilities from the NVD \cite{NVD} that meet the following conditions:
(i) the software repositories on GitHub should involve the vulnerable version and the patched version corresponding to the diff file of the vulnerability, 
because VulTrigger depends on the program analysis of these versions; 
(ii) at least a patch statement of the vulnerability should belong to a function, meaning that the vulnerabilities whose patch statements are all outside of a function are filtered, 
because we focus on inter-procedural vulnerabilities.  
We build the InterPVD dataset for C/C++ open-source software, which involves 769 vulnerabilities belonging to 16 CWEs and 53 software products. 
Each vulnerability is labeled with its patch statements, its vulnerability-triggering statements, its nature in terms of being inter-procedural or not, its type of inter-procedural vulnerability, and its sequence of function calls starting from the to-be-patched function
to the vulnerability-triggering function.

Our data collection process has three steps. First, we collect the vulnerabilities whose diff files can be obtained from the patch links described in the NVD \cite{NVD}. 
Second, we filter the vulnerabilities whose diff files do not satisfy the 
heuristics described in \cite{DBLP:conf/acsac/LiZXJQH16}, and then manually check the remaining ones for their relevance and correctness. 
Third, for each vulnerability that has multiple different commits, we 
decide whether to select one commit or merge multiple commits as follows. (i) If the commits are similar, indicating that they are the patches for different versions, we randomly select one commit. (ii) If the commits are not similar, indicating that they are multiple revisions for the same vulnerability, we merge these commits 
under the condition that the changes described in these commits are in different locations and the context of each commit is the same as the corresponding code in the versions to which the other commits belong; otherwise, we do not consider such a vulnerability because automatically merging these commits is difficult.
We apply VulTrigger to identify vulnerability-triggering statements for the 769 vulnerabilities and manually check their correctness and types by 5 experienced researchers. Each vulnerability is checked by at least two researchers to ensure consistent results. 
During the process of manual check, we found 12 vulnerabilities to which NVD assigned inaccurate CWE IDs and we corrected them. 
We reported this to the NVD team and the team replied by stating that they would revisit it.
Table \ref{Table_wrong_cwe} lists the vulnerabilities in the NVD with incorrect CWE IDs and their corresponding correct CWE IDs.

\begin{table}[!tb]
	\caption{12 vulnerabilities in the NVD with incorrect CWE IDs identified by the present study
 }
	\vspace{-0.4cm}
	\label{Table_wrong_cwe}
	\scriptsize
	\centering
	\begin{tabular}{|c|c|c|c|}
		\hline
		No. & CVE ID & CWE ID in the NVD & Correct CWE ID \\
		\hline
		1 & CVE-2009-1897 & CWE-119 & CWE-476\\
		\hline
		2 & CVE-2009-2767 & CWE-119 & CWE-476\\
		\hline
		3 & CVE-2009-1298 & CWE-119 & CWE-476 \\
		\hline
		4 & CVE-2014-0205 & CWE-119 & CWE-416 \\
		\hline
		5 & CVE-2015-4002 & CWE-119 & CWE-835\\
		\hline
		6 & CVE-2019-13295 & CWE-125 & CWE-369 \\
		\hline
		7 & CVE-2019-13297 & CWE-125 & CWE-369 \\
		\hline
		8 & CVE-2009-4307 & CWE-189 & CWE-369\\
		\hline
		9 & CVE-2013-6367 & CWE-189 & CWE-369\\
		\hline
		10 & CVE-2015-4003 & CWE-189 & CWE-369\\
		\hline
        11 & CVE-2016-2070 & CWE-189 & CWE-369 \\
		\hline
		12 & CVE-2018-5816 & CWE-190 & CWE-369 \\
		\hline
	\end{tabular}
	\vspace{-0.2cm}
\end{table}

\vspace{-0.2cm}
\subsection{Effectiveness of VulTrigger}
\label{subsec:RQ1}
\noindent
{\bf Evaluation metrics.}
To facilitate the comparison, we adapt the standard metrics \cite{DBLP:journals/csur/PendletonGCX17,li2021sysevr} to this setting, including {\em False Positive Rate} (FPR), {\em False Negative Rate} (FNR), accuracy, precision, and F1-measure (F1). We prefer to achieve low FNR, low FPR, high accuracy, high precision, and high F1.

\noindent{\bf Effectiveness on vulnerability-triggering statements.}
To evaluate the effectiveness of VulTrigger in identifying the vulnerability-triggering statements with respect to given vulnerability patches, we use {\em accuracy} to measure the proportion of vulnerabilities whose vulnerability-triggering statements are correctly identified among all vulnerabilities. We consider the vulnerability-triggering statements to be correctly identified if the identified statements include one true vulnerability-triggering statement, while noting that screening the identified vulnerability-triggering statements is left to manual analysis. 
Given that there are no existing methods for identifying vulnerability-triggering statements with respect to a given patch, we 
leverage the side-product capabilities of SAST tools in detecting vulnerability-triggering statements. We stress that these tools 
do not need patches as input because they are designed to detect vulnerabilities, but they have a side-product capability in detecting vulnerability-triggering statements. 
SAST tools can be divided into two categories according to whether one requires to compile programs or not. 
For SAST tools that do not require program compilation, we consider 
Flawfinder \cite{Flawfinder}
and Checkmarx \cite{Checkmarx}.
For SAST tools that require program compilation, we consider Infer \cite{Infer}, CodeQL \cite{CodeQL}, and Fortify \cite{Fortify}.
Since building the compilation environments to accommodate all vulnerabilities incurs a very high labor cost, we only consider 
the vulnerabilities in the Magma dataset \cite{DBLP:journals/pomacs/HazimehHP20}, which 
was originally created for 
evaluating the effectiveness of fuzzers. 
We consider all 
72 vulnerabilities corresponding to the 16 CWEs described in Table \ref{Table_cwe_sink_type} for comparison.
For VulTigger, we use Joern \cite{DBLP:conf/sp/YamaguchiGAR14} to generate the program dependency graph, set threshold $\theta=3$ in the process of generating program slices of critical variables, and remove the vulnerabilities whose programs cannot pass the program analysis of Joern. 

\begin{table}[!tb]
	\caption{Comparing VulTrigger with two SAST tools which do not require program compilation based on 769 vulnerabilities
 }
	\vspace{-0.4cm}
	\label{Table_accuracy_sink}
	\scriptsize
	\centering
	\begin{tabular}{|c|c|c|m{.11\textwidth}<{\centering}|}
		\hline
  Method & {\color{black}\#Supported CWEs} & Accuracy {\color{black}(\%)} & \#Vulnerability-triggering statements\\

		\hline
        Ground truth & {\color{black}16} & 100.0 
        & 1.2\\
		\hline
		Flawfinder \cite{Flawfinder}& {\color{black}9} & 9.8 
  & 14.3\\
		\hline
        Checkmarx \cite{Checkmarx}& {\color{black}16} & 12.7 
        & 114.7\\
		\hline
        VulTrigger & {\color{black}16} &{\bf 69.8}
        & {\bf 2.3}\\
		\hline
	\end{tabular}
	\vspace{-0.2cm}
\end{table}

\begin{table}[!tb]
	\vspace{-0.2cm}
	\caption{Comparing VulTrigger and five SAST tools based on 72 vulnerabilities}
	\vspace{-0.4cm}
	\label{Table_compiling_tools}
	\scriptsize
	\centering
	\begin{tabular}{|c|m{.06\textwidth}<{\centering}|m{.05\textwidth}<{\centering}|m{.07\textwidth}<{\centering}|m{.11\textwidth}<{\centering}|}
		\hline
		Method & Compiling programs? & {\color{black}\#Supported CWEs} & Accuracy {\color{black}(\%)}& \#Vulnerability-triggering statements \\
		\hline
        Ground truth & - & {\color{black}16} & 100.0 
        & 1.1 \\
		\hline		
        Flawfinder \cite{Flawfinder} & No & {\color{black}9} & 5.9 
        & 9.1 \\
		\hline
        Checkmarx \cite{Checkmarx}& No & {\color{black}16} & 31.9 
        & 275.2 \\
		\hline
        Infer \cite{Infer}& Yes & {\color{black}16} & 2.5 
        & 3.3 \\
		\hline
        CodeQL \cite{CodeQL}& Yes & {\color{black}15} & 8.6 
        & 2.8 \\
		\hline
        Fortify \cite{Fortify}& Yes & {\color{black}13} & 13.0 
        & 31.8 \\
		\hline
        VulTrigger & No & {\color{black}16} & {\bf 66.7} 
        & {\bf 1.2} \\
		\hline    
	\end{tabular}
	\vspace{-0.4cm}
\end{table}

Table \ref{Table_accuracy_sink} compares VulTrigger and SAST tools which do not require program compilation via the
769 vulnerabilities {\color{black}corresponding to 16 CWEs. We observe that Flawfinder supports 9 CWEs involving 520 vulnerabilities and the other tools support 16 CWEs.
We evaluate the accuracy of these SAST tools via the vulnerabilities whose CWEs are supported by them.}
We also observe that Flawfinder and Checkmarx achieve a very low accuracy for each CWE, leading to an average accuracy of 11.9\%. 
Table \ref{Table_compiling_tools} compares VulTrigger and SAST tools
via the 
72 vulnerabilities derived from the Magma dataset \cite{DBLP:journals/pomacs/HazimehHP20}.  
We observe that Flawfinder supports 9 CWEs involving 51 vulnerabilities, that CodeQL supports 15 CWEs involving 58 vulnerabilities, that Fortify supports 13 CWEs involving 49 vulnerabilities, and that the other tools supports 16 CWEs.
We further observe that all these SAST tools
achieve a very low accuracy.
This can be understood as follows: (i) SAST tools have high FNRs in detecting vulnerabilities; and (ii) their goals are not to
detect vulnerability-triggering statements of given vulnerabilities.
By contrast, VulTrigger leverages the patch statements and achieves on average a 55.8\% higher accuracy than that of the SAST tools.

To show how many vulnerability-triggering statements which are identified but need to be filtered by human analysis, we compare the average number of true vulnerability-triggering statements and the average number of vulnerability-triggering statements identified by SAST tools and VulTrigger. 
For each vulnerability, we focus on the vulnerability-triggering statements identified 
in the file involving the true vulnerability-triggering function(s).
From Table \ref{Table_accuracy_sink} and Table \ref{Table_compiling_tools}, we observe that 
VulTrigger needs human experts to filter 1.1 (0.1) statements for 769 (72) vulnerabilities on average. 
Checkmarx
demands human experts to filter, on average, 113.5 (274.1) statements for each of 769 (72) vulnerabilities, 
achieving a higher accuracy 
at the cost of filtering much more statements by human experts; the other SAST tools 
demand human experts to filter on average 13.1 (10.7) statements. 
This can be explained by the fact that SAST tools report many false vulnerable statements.

\noindent{\bf Comparing inter-procedural vulnerability identification methods.} 
Existing studies related to inter-procedural vulnerabilities \cite{DBLP:conf/ccs/LiP17,DBLP:conf/icse/LiuMZGLLSH020,DBLP:conf/cns/WangW0BJ20} (dubbed ``patch-function method'') focus on whether or not multiple functions are involved in a vulnerability patch. 
That is, these studies consider a vulnerability inter-procedural as long as its patch involves multiple functions. 
As shown in Table \ref{Table_compare_with_existing_method}, 
VulTrigger can improve FPR by 20.6\%, FNR by 32.7\%, accuracy by 22.8\%, precision by 66.4\%, and F1 by 45.1\% in identifying inter-procedural vulnerabilities. 
Nevertheless, VulTrigger overlooks some vulnerability-triggering statements,  
which needs to be addressed in future work. 
We manually analyze these false negatives and summarize our findings as follows.
(i) VulTrigger cannot identify implicit semantic associations between variables (e.g., CVE-2014-9664), which causes incorrect transformations of critical variables. 
(ii) If the diff file  (e.g., CVE-2014-9604) only adds checking statements that are not data-dependent on critical variables, the resulting program slice cannot involve the vulnerability-triggering statement, causing the vulnerability-triggering statement to be missed. 
(iii) For some vulnerabilities (e.g., CVE-2015-5706), false negatives occur because there are no critical variables or the critical variables cannot be extracted correctly in the diff files. 
(iv) Some inter-procedural vulnerabilities are missed or incorrectly identified because some complex statements are missed or parsed incorrectly by Joern \cite{DBLP:conf/sp/YamaguchiGAR14}, leading to wrong vulnerability-triggering statements.

\begin{table}[!tb]
	\caption{Comparing VulTrigger and the existing methods that have some capabilities in detecting inter-procedural vulnerabilities (unit: \%)
	}
	\vspace{-0.4cm}
	\label{Table_compare_with_existing_method}
	\scriptsize
	\centering
	\begin{tabular}{|c|c|c|c|c|c|c|}
		\hline
		Method & FPR & FNR & Accuracy & Precision & F1 \\
		\hline
		Patch-function \cite{DBLP:conf/ccs/LiP17,DBLP:conf/icse/LiuMZGLLSH020,DBLP:conf/cns/WangW0BJ20}& 23.8 & 90.9 & 64.0 & 7.8 & 8.4 \\
		\hline
		VulTrigger & {\bf 3.2} & {\bf 58.2} & {\bf 86.8} & {\bf 74.2} & {\bf 53.5}\\
		\hline
	\end{tabular}
	\vspace{-0.6cm}
\end{table}

\begin{insight}
{VulTrigger substantially outperforms the side-product capabilities of existing SAST tools in identifying vulnerability-triggering statements (while noting that there are no known methods specifically designed to identify vulnerability-triggering statements), by achieving a 55.8\% higher accuracy on average in identifying vulnerability-triggering statements, and by improving FPR by 20.6\%, FNR by 32.7\%, and F1 by 45.1\% in identifying inter-procedural vulnerabilities. }
\end{insight}

\subsection{Distribution of Inter-Procedural Vulnerabilities}
\label{subsec:RQ2}
\begin{figure}[!tb]
	\centering
	\includegraphics[width=0.3\textwidth]{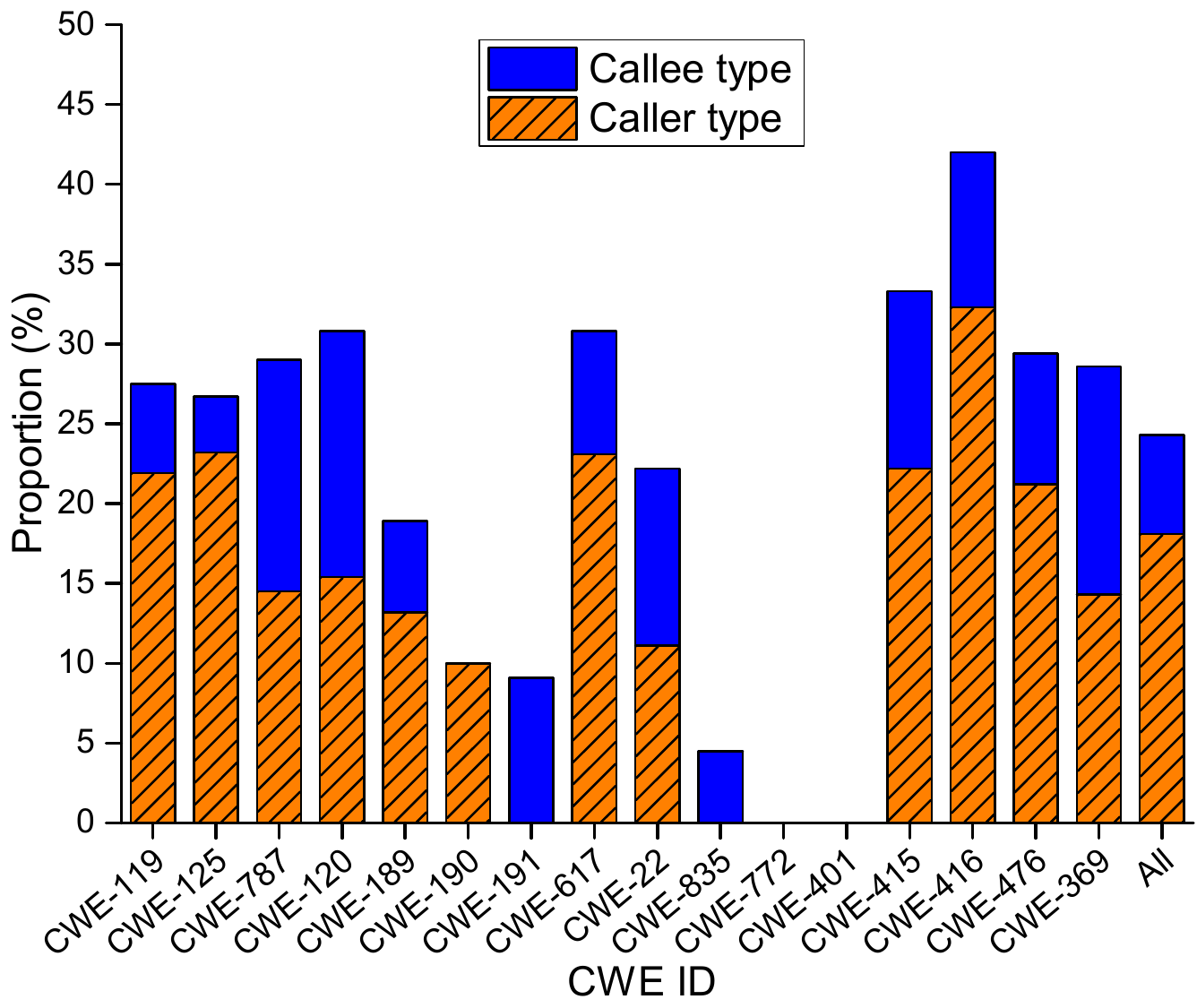}
	\vspace{-0.4cm}
	\caption{Illustrating the proportion of different types of inter-procedural vulnerabilities corresponding to each CWE}
	\vspace{-0.4cm}
	\label{Fig_cross_function_cwe}
\end{figure}

{\noindent \bf Proportion of inter-procedural vulnerabilities.}
To see what proportion of different types of inter-procedural vulnerabilities, we consider all the 16 CWEs in our InterPVD dataset. 
Among them, CWE-119, CWE-125, and CWE-476 are the top-3 CWEs involving 57.5\% vulnerabilities and 65.2\% inter-procedural vulnerabilities. 
Figure \ref{Fig_cross_function_cwe} depicts the proportion of different types of inter-procedural vulnerabilities in different CWEs. 
We observe that 24.3\% vulnerabilities are inter-procedural vulnerabilities, including 18.1\% of the caller type and 6.2\% of the callee type.
We also observe that there are no inter-procedural vulnerabilities for CWE-772 and CWE-401. This can be explained by the fact that the types of vulnerability-triggering statements for CWE-772 and CWE-401 are usually the {\tt return} statement or the last statement of the to-be-patched function (i.e., T-10 and T-11), where the vulnerability-triggering statements appear in the scope of the to-be-patched function. 
For CWE-191 and CWE-835, there are only inter-procedural vulnerabilities of the callee type. We speculate that this is caused by the small number of vulnerabilities of CWE-191 and CWE-835 in our dataset, involving 11 vulnerabilities of CWE-191 and 22 vulnerabilities of CWE-835.
For most CWEs, we have more inter-procedural vulnerabilities of the caller type, indicating that more vulnerabilities are patched in the direct or indirect calling functions.  

{\noindent \bf Number of inter-procedural layers.} 
To show how many functions are involved in inter-procedural vulnerabilities, we summarize the number of inter-procedural layers from the vulnerability patch function to the vulnerability-triggering function. 
If a vulnerability has multiple pairs mapping from the vulnerability patch function to the vulnerability-triggering function, we take them as multiple vulnerability instances. We obtain the average number of inter-procedural layers starting at the vulnerability patch function and ending at the vulnerability-triggering function for all vulnerability instances.
Table \ref{Table_num_cross_function_cwe} 
summarizes the average number of layers of inter-procedural vulnerabilities for each CWE. 
For each CWE 
that has both types of inter-procedural vulnerabilities, the inter-procedural vulnerabilities of the callee type have more layers than those of the caller type. This indicates that vulnerability-triggering statements appear more often in the functions that indirectly call the to-be-patched function.
The average number of layers of inter-procedural vulnerabilities is 2.8, and the inter-procedural vulnerabilities of the callee type has 0.9 layers more than those of the caller type.

\begin{table*}[!htbp]
	\caption{Average number of layers of different types of inter-procedural vulnerabilities for 16 CWEs}
	\vspace{-0.4cm}
	\label{Table_num_cross_function_cwe}
	\scriptsize
	\centering
	\begin{tabular}{|c|c|c|c|c|c|c|c|c|c|c|c|c|c|c|c|c||c|}
		\hline
		\diagbox{Type}{CWE ID} & 119 & 125 & 787 & 120  & 189 & 190 & 191 & 617 & 22 & 835 & 772 & 401  & 415 & 416 & 476 & 369 & All\\
		\hline
        Caller type & 2.7 & 2.2 & 2.3 & 3.0 & 2.6 & 3.2 & - & 3.3 & 2.0 & - & - & - & 2.0 & 2.5 & 2.7 & 2.7 & 2.6\\
		\hline
        Callee type & 3.3 & 2.6 & 2.6 & 4.0 & 3.0 & - & 3.0 & 6.0 & 4.0 & 3.0 & - & - & 3.0 & 3.0 & 3.9 & 7.7 & 3.5 \\
		\hline
        \hline
		All & 2.8 & 2.3 & 2.4 & 3.5 & 2.7 & 3.2 & 3.0 & 4.0 & 3.0 & 3.0 & - & - & 2.3 & 2.6 & 3.0 & 5.2 & 2.8\\
		\hline
	\end{tabular}
	\vspace{-0.2cm}
\end{table*}

\begin{insight}
{Among 769 vulnerabilities, 24.3\% are inter-procedural vulnerabilities with 2.8 layers on average, including 18.1\% of the caller type and 6.2\% of the callee type.}
\end{insight}

%% file: Evaluation.tex
\noindent
\noindent{\bf Experimental setup.}
{\color{black}To evaluate the effectiveness of function-level vulnerability detectors (trained by existing datasets) in detecting intra-procedural vulnerabilities and inter-procedural vulnerabilities,}
we consider five function-level vulnerability detectors \cite{DBLP:conf/ijcnn/HanifM22,DBLP:conf/msr/FuT22,DBLP:conf/nips/ZhouLSD019,DBLP:journals/tse/ChakrabortyKDR22,DBLP:conf/icse/WuZD0X022} because they represent the state-of-the-art: VulBERTa \cite{DBLP:conf/ijcnn/HanifM22} and LineVul \cite{DBLP:conf/msr/FuT22} 
use Transformer-based pre-training model;
Devign \cite{DBLP:conf/nips/ZhouLSD019} 
and {\scshape ReVeal} \cite{DBLP:journals/tse/ChakrabortyKDR22} identify vulnerabilities by  Graph Neural Network-based models;
VulCNN \cite{DBLP:conf/icse/WuZD0X022} converts the graph representation of code to an image 
while leveraging Convolutional Neural Network-based models.

For the training set, we use the {\scshape ReVeal} dataset \cite{DBLP:journals/tse/ChakrabortyKDR22} because it also represents the state-of-the-art.
The {\scshape ReVeal} dataset labels the to-be-patched functions as vulnerable, using the same labeling scheme as other datasets. 
For each of aforementioned five methods, we apply 10-fold cross-validation to train the model. 
For test set, we use the InterPVD dataset 
in which vulnerabilities can be divided into {\em inter-procedural} and {\em intra-procedural} vulnerability test sets.
The inter-procedural vulnerability test set can be further divided into {\em caller} and {\em callee} test sets based on whether the inter-procedural vulnerabilities are caller or callee types. 

The samples in the test set contains vulnerability-triggering functions, to-be-patched functions, and their corresponding patched functions of vulnerabilities, which are labeled as vulnerable or non-vulnerable.
We conduct two experimental tasks: detecting to-be-patched functions vs. vulnerability-triggering functions. In the former case,
the to-be-patched functions are considered vulnerable but the patched functions and vulnerability-triggering functions are considered non-vulnerable. When detecting vulnerability-triggering functions, vulnerability-triggering functions are considered vulnerable but the patched functions and to-be-patched functions are considered non-vulnerable.

\noindent{\bf Experimental results.} 
To assure fair comparison, we set FNR to 20.0\% for each detector by adjusting the probability threshold when determining whether a piece of code is vulnerable or not. This value is chosen because it corresponds to the FNR 
of the model that achieves the highest F1. 
Table \ref{Table_effectiveness_all} summarizes the experimental results via the InterPVD test set. 
We observe that there is little variation in each evaluation metric of the five detectors in detecting to-be-patched functions and vulnerability-triggering functions. We also observe that the five detectors show significantly lower effectiveness in all metrics, except FNR, in detecting vulnerability-triggering functions when compared with detecting to-be-patched functions. The overall effectiveness, F1 score, for detecting vulnerability-triggering functions is, on average, 18.7\% lower than that of detecting to-be-patched functions. 
This can be explained by the fact that to-be-patched functions are labeled as vulnerable in the training set, meaning that the detectors trained from this dataset tend to have lower effectiveness in detecting vulnerability-triggering functions in the test set.

\begin{table}[!tbp]
	\caption{Effectiveness comparison via InterPVD (unit: \%)}
	\vspace{-0.4cm}
	\label{Table_effectiveness_all}
	\scriptsize
	\centering
	\begin{tabular}{|c|c|c|c|c|c|}
		\hline
		Method & FPR & FNR & Accuracy & Precision &  F1 \\
		\hline
        \multicolumn{6}{|c|}{Detecting to-be-patched functions} \\
        \hline
		VulBERTa & 80.1 & 20.0 & 47.7 & 46.2 & 58.6 \\
		\hline
		LineVul & 77.8 & 20.0 & 48.9 & 46.9 & 59.1 \\
		\hline
        Devign & 81.8 & 20.0 & 46.5 & 45.3 & 57.8 \\
		\hline
		{\scshape ReVeal} & 76.0 & 20.0 & 49.7 & 47.1 &  59.3 \\
		\hline
        VulCNN & 80.3 & 20.0 & 47.2 & 45.6 & 58.1 \\
		\hline
        \multicolumn{6}{|c|}{Detecting vulnerability-triggering functions} \\
        \hline
		VulBERTa & 79.9 & 20.0 & 35.2 & 25.3 & 38.5 \\
		\hline
		LineVul & 78.5 & 20.0 & 36.3 & 25.6 & 38.8 \\
		\hline
        Devign & 82.3 & 20.0 & 35.1 & 27.3 & 40.7 \\
		\hline
		{\scshape ReVeal} & 82.3 & 20.0 & 35.1 & 27.3 & 40.7 \\
		\hline
        VulCNN & 80.6 & 20.0 & 36.0 & 27.3 & 40.7 \\
		\hline
	\end{tabular}
	\vspace{-0.4cm}
\end{table}

Table \ref{Table_effectiveness_inter_intra} summarizes the experimental results based on the inter-procedural and intra-procedural vulnerability test sets. 
We observe that the effectiveness in detecting to-be-patched functions is higher than detecting vulnerability-triggering functions.
This can be attributed to the fact that existing vulnerability datasets only label the to-be-patched functions as vulnerable, 
implying limited capability in learning features associated with vulnerability-triggering functions.
In terms of detecting to-be-patched functions, the effectiveness with respect to inter-procedural vulnerabilities is consistently lower than its counterpart with respect to intra-procedural vulnerabilities. Both intra-procedural and inter-procedural vulnerabilities have some to-be-patched functions, which are labeled as vulnerable in the Reveal dataset, meaning no bias favoring detection of to-be-patched functions. 
Experimental results indicate that detecting to-be-patched functions of inter-procedural vulnerabilities is more challenging than detecting their counterparts of intra-procedural vulnerabilities.

For inter-procedural vulnerabilities, Table \ref{Table_effectiveness_caller_callee} compares the five detectors based on the caller and callee test sets. 
We observe that vulnerability detectors achieve, on average, a 4.6\% higher overall effectiveness F1 in detecting to-be-patched functions for the caller test set than the callee test set, and a 7.2\% lower F1 in detecting vulnerability-triggering functions for the caller test set than the callee test set. 
This could be attributed to the increased ratio of non-vulnerable to vulnerable samples in the callee test set, which leads to more false positives and significantly lowers the overall effectiveness. 
Specifically, for detecting to-be-patched functions, the ratio of non-vulnerable to vulnerable samples is 1.5:1 in the callee test set and 1.6:1 in the caller test set; for detecting vulnerability-triggering functions, the ratio is 4.5:1 in the callee test set and 3.9:1 in the caller test set. 

\begin{insight}
{
Function-level vulnerability detectors are much less effective in detecting vulnerability-triggering functions than detecting to-be-patched functions; detecting to-be-patched functions of inter-procedural vulnerabilities is more challenging than detecting their counterparts of intra-procedural vulnerabilities.
}
\end{insight}

\begin{table}[!tbp]
	\caption{Effectiveness  of five detectors via inter-procedural and intra-procedural vulnerability test sets (unit: \%)}
	\vspace{-0.4cm}
	\label{Table_effectiveness_inter_intra}
	\scriptsize
	\centering
	\begin{tabular}{|c|c|c|c|c|c|c|}
		\hline
		Method & Test set & FPR & FNR & Accuracy & Precision &  F1 \\
		\hline
        \multicolumn{7}{|c|}{Detecting to-be-patched functions} \\
        \hline
		\multirow{2}{*}{VulBERTa} & Inter-procedural & 75.8 & 23.8 & 45.5 & 41.1 & 53.4 \\
        \cline{2-7}
        & Intra-procedural & 81.9 & 18.8 & 48.5 & 47.9 & 60.3 \\
		\hline
		\multirow{2}{*}{LineVul} & Inter-procedural & 75.2 & 18.7 & 47.9 & 42.8 & 56.1 \\
        \cline{2-7}
        & Intra-procedural & 78.9 & 20.4 & 49.3 & 48.4 & 60.2\\
		\hline
        \multirow{2}{*}{Devign} & Inter-procedural & 84.3 & 20.3 & 41.7 & 39.2 & 52.6 \\
        \cline{2-7}
        & Intra-procedural & 80.7 & 19.9 & 48.5 & 47.9 & 59.9\\
		\hline
		\multirow{2}{*}{{\scshape ReVeal}} & Inter-procedural & 72.9 & 21.2 & 48.1 & 42.5 & 55.3 \\
        \cline{2-7}
        & Intra-procedural & 77.5 & 19.6 & 50.3 & 49.0 & 60.9\\		
         \hline
        \multirow{2}{*}{VulCNN} & Inter-procedural & 73.7 & 25.3 & 46.0 & 40.9 & 52.9 \\
		\cline{2-7}
        & Intra-procedural & 83.6 & 18.1 & 47.8 & 47.4 & 60.0\\
        \hline
        \multicolumn{7}{|c|}{Detecting vulnerability-triggering functions} \\
        \hline
		\multirow{2}{*}{VulBERTa} & Inter-procedural & 77.7 & 30.8 & 31.9 & 18.6 & 29.3 \\
        \cline{2-7}
        & Intra-procedural & 80.9 & 17.0 & 36.5 & 27.7 & 41.5 \\
		\hline
		\multirow{2}{*}{LineVul} & Inter-procedural & 81.2 & 35.9 & 28.1 & 16.8 & 26.6 \\
        \cline{2-7}
        & Intra-procedural & 77.4 & 15.8 & 39.3 & 28.8 & 42.9\\
		\hline
        \multirow{2}{*}{Devign} & Inter-procedural & 81.6 & 9.5 & 34.0 & 23.5 & 37.3 \\
        \cline{2-7}
        & Intra-procedural & 82.6 & 23.0 & 35.5 & 28.9 & 42.1\\
		\hline
		\multirow{2}{*}{{\scshape ReVeal}} & Inter-procedural & 81.1 & 23.3 & 31.5 & 20.8 & 32.7 \\
        \cline{2-7}
        & Intra-procedural & 82.9 & 18.9 & 36.6 & 29.9 & 43.7\\		
         \hline
        \multirow{2}{*}{VulCNN} & Inter-procedural & 74.1 & 25.9 & 36.2 & 21.3 & 33.1 \\
		\cline{2-7}
        & Intra-procedural & 83.7 & 18.2 & 35.9 & 29.5 & 43.4\\
        \hline
	\end{tabular}
	\vspace{-0.2cm}
\end{table}

\begin{table}[!tbp]
	\caption{Effectiveness comparison on the caller and callee test sets of inter-procedural vulnerabilities (unit: \%)}
	\vspace{-0.4cm}
	\label{Table_effectiveness_caller_callee}
	\scriptsize
	\centering
	\begin{tabular}{|c|c|c|c|c|c|c|}
		\hline
		Method & Test set & FPR & FNR & Accuracy & Precision &  F1 \\
		\hline
        \multicolumn{7}{|c|}{Detecting to-be-patched functions} \\
        \hline
		\multirow{2}{*}{VulBERTa} & Caller & 77.2 & 20.2 & 46.3 & 42.0 & 55.0\\
        \cline{2-7}
        & Callee & 69.8 & 40.5 & 41.9 & 36.2 & 45.1 \\
		\hline
		\multirow{2}{*}{LineVul} & Caller & 75.7 & 16.1 & 48.8 & 43.7 & 57.5 \\
        \cline{2-7}
        & Callee & 73.0 & 30.9 & 43.8 & 38.7 & 49.6\\
		\hline
        \multirow{2}{*}{Devign} & Caller & 83.3 & 21.4 & 42.0 & 39.5 & 52.6 \\
        \cline{2-7}
        & Callee & 89.1 & 14.3 & 40.0 & 38.0 & 52.6\\
		\hline
		\multirow{2}{*}{{\scshape ReVeal}} & Caller & 76.0 & 19.2 & 47.3 & 42.5 & 55.7 \\
        \cline{2-7}
        & Callee & 58.2 & 31.4 & 52.2 & 42.9 & 52.8\\		
         \hline
        \multirow{2}{*}{VulCNN} & Caller & 74.1 & 24.6 & 46.1 & 41.2 & 53.3 \\
		\cline{2-7}
        & Callee & 71.9 & 28.9 & 45.3 & 39.7 & 50.9\\
        \hline
        \multicolumn{7}{|c|}{Detecting vulnerability-triggering functions} \\
        \hline
		\multirow{2}{*}{VulBERTa} & Caller & 80.3 & 30.1 & 29.6 & 17.7 & 28.3\\
        \cline{2-7}
        & Callee & 65.4 & 33.3 & 41.9 & 23.2 & 34.4\\
		\hline
		\multirow{2}{*}{LineVul} & Caller & 83.8 & 39.8 & 25.0 & 15.1 & 24.1 \\
        \cline{2-7}
        & Callee & 69.1 & 20.8 & 41.9 & 25.3 & 38.4\\
		\hline
        \multirow{2}{*}{Devign} & Caller & 80.7 & 9.8 & 33.9 & 22.6 & 36.1 \\
        \cline{2-7}
        & Callee & 86.4 & 8.3 & 34.4 & 27.9 & 42.7\\
		\hline
		\multirow{2}{*}{{\scshape ReVeal}} & Caller & 82.4 & 20.6 & 30.4 & 20.1 &  32.1\\
        \cline{2-7}
        & Callee & 74.2 & 33.3 & 36.7 & 24.6 & 36.0\\		
         \hline
        \multirow{2}{*}{VulCNN} & Caller & 74.7 & 25.8 & 35.4 & 20.6 & 32.2 \\
		\cline{2-7}
        & Callee & 70.8 & 26.1 & 40.0 & 25.0 & 37.4\\
        \hline
	\end{tabular}
	\vspace{-0.4cm}
\end{table}

%% file: Discussion.tex
\smallskip
\noindent{\bf Use cases of
VulTrigger}.
Two example use cases of VulTrigger are the following.
First, VulTrigger can be used to 
label a large number of vulnerability-triggering functions, 
which can be leveraged to improve the effectiveness of function-level vulnerability detectors in coping with vulnerability-triggering functions.
Second, VulTrigger can also be used to extract more accurate statements related to inter-procedural vulnerabilities, leading to high-quality program slices. These high-quality program slices, along with their graph representations, can be used as input examples to train more effective slice-level vulnerability detectors in detecting inter-procedural vulnerabilities  (e.g., \cite{VulChecker}).

\noindent{\bf Threats to validity.} 
There are two threats to the validity of our study. First, we use manual analysis to establish the ground truth of vulnerability-triggering statements in C/C++ open-source software. 
Although each vulnerability is checked by at least two researchers,
it is still possible that we miss some vulnerability-triggering statements because our manual check may still miss some execution paths. This is true because identifying the complete set of execution paths and thus the complete set of vulnerability-triggering statements is difficult even by manual analysis. To mitigate the threat, a more capable static analysis tool is needed to accurately identify all execution paths.  
Second, we create and maintain a list of keywords of APIs/system calls, 
which may trigger out-of-bounds access or memory errors. If a user-defined function name contains a keyword in the above list and directly or indirectly makes a system call, the user-defined function call is considered the vulnerability-triggering statement. This
may reduce the number of inter-procedural layers.

\noindent{\bf Limitations.} 
The present study has four limitations. First, we focus on inter-procedural vulnerabilities in C/C++ open-source software. Future studies need to consider other programming languages.
Second, we only consider vulnerability-triggering statements related to 16 CWEs. Future studies need to investigate more 
CWEs. 
Third, we identify 
as many vulnerability-triggering statements as possible for a given vulnerability. Future studies need to consider how to identify all vulnerability-triggering statements in all execution paths with respect to a vulnerability.
Fourth, future studies need to reduce false negatives of VulTrigger 
as discussed in Section \ref{subsec:RQ1} and propose effective inter-procedural vulnerability detection approaches.

%% file: RelatedWork.tex
\noindent{\bf Prior studies on analyzing open-source software vulnerabilities.}
These studies are from various perspectives, such as: (i) {\em distributions} of project-level vulnerabilities \cite{DBLP:conf/icse/LiuMZGLLSH020}, of vulnerabilities in crowd-sourced software \cite{9195034}, and of vulnerabilities in cloned vs. non-cloned code \cite{DBLP:journals/compsec/KimL18, DBLP:conf/esem/IslamZN17}; (ii) {\em patches} \cite{DBLP:conf/ccs/LiP17, DBLP:conf/dsn/Wang0BJ19}, patch explanations \cite{DBLP:conf/issre/LiangHZ0X019}, and security impact of patches \cite{DBLP:conf/ndss/WuHML20}; 
(iii) {\em characteristics} of project dependencies related to vulnerabilities \cite{DBLP:conf/dsn/LiSTWL21}, of features indicating vulnerabilities \cite{DBLP:journals/tifs/ZhangCWR19}, and of vulnerable code changes \cite{DBLP:conf/sigsoft/BosuCHHJ14}.
To our knowledge, we are the first to systematically study inter-procedural vulnerabilities of open-source software, despite the previous studies \cite{DBLP:conf/ccs/LiP17,DBLP:conf/icse/LiuMZGLLSH020,DBLP:conf/cns/WangW0BJ20} that investigate whether or not multiple functions are involved in a vulnerability patch and cannot accurately identify inter-procedural vulnerabilities 
(Table \ref{Table_compare_with_existing_method}).

\noindent{\bf Prior studies on identifying 
{\em to-be-patched} 
vs. {\em vulnerability-triggering} statements}. 
(i) To-be-patched statements 
are the 
root cause of a vulnerability \cite{DBLP:conf/issta/LippBP22} 
and can be identified by various methods, such as clone-based detection \cite{DBLP:conf/acsac/LiZXJQH16,DBLP:conf/sp/JangAB12}, deep learning-based detection \cite{DBLP:journals/tdsc/LiZXCZ022}, statistical localization enabled by exploits \cite{DBLP:conf/asiaccs/ShenKDSR21}. 
(ii) Vulnerability-triggering statements 
can be identified by static and dynamic methods.
Static methods are demonstrated by SAST tools \cite{Flawfinder,CodeQL,Infer,Checkmarx,Fortify}, 
which have high false-positive rates and high false-negative rates. For example, the best static analysis tool misses 
47\%-80\% vulnerabilities 
\cite{DBLP:conf/issta/LippBP22}, which is also confirmed by 
our experiments (cf. Tables \ref{Table_accuracy_sink} and \ref{Table_compiling_tools}).
VulTrigger is a static method to identify vulnerability-triggering statements.
Dynamic methods (e.g., fuzzing \cite{DBLP:journals/tse/ManesHHCESW21,DBLP:conf/sp/GanZQTLPC18,DBLP:conf/icse/MengDLBR22}) can identify vulnerability-triggering statements 
but incur high false-negative rates because it is difficult to test all execution paths.
Moreover, they incur a very high overhead 
when attempting to trigger a vulnerability (i.e., not scalable).

\noindent{\bf Prior studies on detecting intra/inter-procedural vulnerabilities.}
{\em Intra-procedural} vulnerability detection methods often take to-be-patched functions and their patched versions to train a detector, while
leveraging machine learning \cite{DBLP:conf/acsac/YamaguchiLR12,DBLP:conf/uss/YamaguchiLR11a} or deep learning \cite{DBLP:conf/ijcai/DuanWJRLYW19,DBLP:conf/nips/ZhouLSD019,DBLP:journals/tifs/WangYTTHFFBW21,DBLP:conf/sigsoft/Li0N21,DBLP:conf/icse/WuZD0X022,DBLP:conf/ijcnn/HanifM22,DBLP:conf/msr/FuT22,DBLP:journals/tse/ChakrabortyKDR22}.
{\em Inter-procedural} vulnerability detection methods can detect vulnerabilities whose to-be-patched statements and vulnerability-triggering statements belong to different functions. 
These methods extract inter-procedural program slices to train a deep learning model \cite{li2021sysevr,DBLP:conf/ndss/LiZXO0WDZ18,VulChecker} or leverage inter-procedural analyses to detect different types of vulnerabilities  \cite{DBLP:conf/ccs/Luo0022,CodeQL,Infer,Checkmarx,Fortify}.
Whereas, VulTrigger determines whether a vulnerability is inter-procedural or not, which allows us to present the first study on the effectiveness of function-level vulnerability detectors in detecting inter-procedural vulnerabilities.